\documentclass[prl,twoside,preprintnumbers,superscriptaddress,twocolumn,nofootinbib,nobibnotes]{revtex4}
\usepackage{amsmath,graphicx,slashed,multirow,color,ulem,setup}
\usepackage{graphicx,epsfig,amssymb,bm,slashed,wasysym,multirow}
\usepackage[utf8]{inputenc}

\preprint{NIKHEF 2019-011}

\begin{document}
\title{Precise predictions for multi-$\TeV$ and $\PeV$ energy neutrino scattering rates}

\author{Rhorry Gauld}
\email{r.gauld@nikhef.nl}
\affiliation{Nikhef, Science Park 105, NL-1098 XG Amsterdam, The Netherlands}

\date{\today}

\begin{abstract}
The scattering rate of multi-$\TeV$ and $\PeV$ energy neutrinos is fast becoming an interesting topic in (astro)particle-physics. This is due to experimental progress at Neutrino Telescopes such as IceCube which have begun to gain sensitivity to the flux of neutrinos in this energy range. In view of this, a precise calculation of the scattering rate of neutrinos upon atoms is presented.
The two main components of the calculation are the differential cross-section predictions for neutrino scattering upon an atomic nucleus (such as that which make up water), as well as upon atomic electrons.
In the first case, the predictions for neutrino-nucleon cross-sections in charged- and neutral-current scattering are refined by including resonant contributions generated within the photon field of the nucleus, which alter the considered distributions by up to $\approx 4\%$.
In the latter case, radiative corrections are provided for all $2\to2$ scattering processes of the form $\bar{\nu}_e  e^-\to f\bar{f}^{\prime}$. For antineutrino energies of $E_{\bar{\nu}_e}\approx6~\PeV$, where these processes become resonantly enhanced (the Glashow resonance) and dominate the total cross-section, these corrections amount to $\approx-10\%$.

\end{abstract}

\maketitle

\section{Introduction}

The measurement of a flux of ultra-high-energy (UHE) neutrinos\footnote{Unless specifically stated, `neutrinos' will collectively refer to both neutrinos and antineutrinos.} at detectors on Earth is extremely important for understanding potential sources of cosmic ray accelerators in our universe. As neutrinos are weakly interacting, they propagate through the Universe without being deflected by magnetic fields or scattering on background photon radiation. 
The detection of these neutrinos within large-volume detectors such as IceCube~\cite{IceCube:2018cha} therefore provides information on their source of production, which in turn can help to identify the source(s) of UHE cosmic rays in our Universe which are expected to generate such a flux.
The potential of this physics program has been recently demonstrated with the identification of the Blazar {\sc \small TXS 0506+056} as a source of high energy neutrinos~\cite{IceCube:2018dnn,IceCube:2018cha}.

The IceCube experiment has already become sensitive to the flux of neutrinos with multi-$\TeV$ and $\PeV$ energies~\cite{Aartsen:2017nbu}, which has enabled various measurements of the cross-sections and event characteristics induced by neutrinos within this energy range~\cite{Aartsen:2017kpd,Bustamante:2017xuy,Aartsen:2018vez}.
A measurement of the event rate of $\PeV$-energy neutrinos is of particular interest as the scattering process of electron antineutrinos upon atomic electrons becomes resonantly enhanced for centre-of-mass (CoM) energies of $\sqrt{S} \approx \sqrt{2 m_e E_{\bar{\nu}_e}} \approx m_W$ (the Glashow resonance~\cite{Glashow:1960zz}), which is achieved for $E_{\bar{\nu}_e} \approx 6~\PeV$. As this process is directly sensitive to the flux of electron antineutrinos at Earth, its measurement provides flavour separation which is invaluable for understanding the production mechanisms which source UHE neutrinos~\cite{Anchordoqui:2004eb, Hummer:2010ai, Xing:2011zm, Biehl:2016psj}.
As more data is collected at IceCube, as well as other large volume detectors such as KM3NeT~\cite{Adrian-Martinez:2016fdl} and Baikal-GVD~\cite{Avrorin:2018ijk}, it is anticipated that a measurement of this process will become feasible.

In anticipation of this experimental progress, it is the purpose of this work to re-visit the corresponding theoretical predictions for the scattering rates of multi-$\TeV$ and $\PeV$ energy neutrinos on atomic targets. Technical details of the calculation are given in the following Section, before applying the calculations to study the following observables: the total inclusive cross-section; the mean inelasticity distribution in muon production; as well as the inclusive cross-section for quark production in $\bar{\nu}_e+e^{-}$ collisions (with focus on the resonance region).

\section{Details of calculation}
The computation of the scattering rate of a high energy neutrino with a stationary atom is performed by separately considering the processes where the neutrino scatters either upon an atomic nucleus or electron, the details of which are summarised below. \\[1mm]
\noindent \textbf{Neutrino-nucleon scattering.}
In the case of a nuclear target, the dominant contributions to the cross-section arise from either the charged-current (CC) or neutral (NC) processes, where an interaction between the incident neutrino and the constituents (partons) of the (potentially bound) target nucleons is mediated by either $\PW$- or $\PZ$-boson exchange respectively. 
The differential cross-section for this process can be conveniently written in terms of the Deep Inelastic Structure (DIS) functions $F_i$ and kinematic variables which characterise the scattering process (see for example Eq.~(2.2) and~(2.3) of Ref.~\cite{Bertone:2018dse}). 
These $F_i$ are a function of the (negative) squared four-momentum $Q^2$ of the exchanged gauge-boson and the variable $x$ (which in the parton model of the nucleus, corresponds to the fraction of the nucleon's momentum carried by the struck parton). They describe the underlying dynamics of the nucleus as probed by the exchanged gauge-boson, and consequently their description is process dependent with respect to the type of interaction (CC/NC), projectile ($\nu/\bar{\nu}$), and target nucleon (bound/free proton/neutron). A prediction for the relevant $F_i$ can be performed perturbatively by convoluting process dependent coefficient functions with a set of parton distribution functions (PDFs) for the (bound) target nucleon. The resultant predictions for the cross-section provided in this way can be made differential with respect to the outgoing lepton kinematics.

The CC and NC processes describe the class of partonic subprocesses where the gauge-boson is exchanged in the $t$-channel, and the four-momentum of the exchanged gauge-boson is time-like. 
In addition to this, there are also partonic subprocesses generated within the photon field of the nucleon which take the form $\bar{\nu}_l \gamma \to f \bar{f}^{\prime} \ell$. These subprocesses (and charge conjugate versions) receive a resonant enhancement when $m_{f\bar{f}^{\prime}} \sim m_W$ (on-shell $\PW$-boson exchange), and additionally receive an enhancement due to collinear splittings of the form $\gamma \to \ell\bar{\ell}$ (on-shell lepton exchange)~\cite{Seckel:1997kk}.
While the contribution of these subprocesses is (naively) suppressed by $\mathcal{O}(\alpha^2)$ as compared to leading-order (LO) CC and NC predictions, they can still impact the total cross-section by several percent~\cite{Alikhanov:2015kla} and should therefore be included. 

To consistently take into account all contributions discussed above, PDF sets based on the NNLO NNPDF3.1luxQED analysis~\cite{Bertone:2017bme} (which uses the luxQED formalism~\cite{Manohar:2016nzj,Manohar:2017eqh}) are generated with the program {\sc\small APFEL}~\cite{Bertone:2013vaa}. The proton PDF set is modified at the input scale $Q_0 = 1.65~\GeV$ to include non-zero (anti)electron and (anti)muon PDFs for the proton according to the fixed-order ansatz assumed in Ref.~\cite{Bertone:2015lqa} (see Eq.~2.6). For the neutron, instead the PDFs are obtained from that of proton PDFs (with zero photon and lepton PDFs) at the scale $Q_0 = 1.65~\GeV$ using isospin symmetry. These PDF sets are then evolved including both NNLO QCD and complete LO QED corrections to the DGLAP equations~\cite{Bertone:2015lqa}.

The predictions for the $F_i$ are obtained by convoluting the resultant PDFs with coefficient functions computed at the same order. Heavy quark mass effects (via FONLL-B,C schemes~\cite{Forte:2010ta,Ball:2015tna}) and nuclear corrections (from the EPPS16 analysis~\cite{Eskola:2016oht,Dulat:2015mca}) are included for the CC and NC predictions as in Ref.~\cite{Bertone:2018dse}.

The computation of the resonantly enhanced contributions is straightforward in this setup, and is obtained by convoluting the evolved lepton PDFs with a partonic cross-section $\bar{\nu}_l \ell \to f \bar{f}^{\prime}$. This approximate calculation includes an all-orders resummation of the collinear logarithms (which dominate the cross-section) via the PDF evolution. In addition, the $\nu\gamma$-induced channel, which is the dominant contribution of the $\mathcal{O}(\alpha)$ correction, is included.
In this approach, the inelastic photon component of the target nucleon (as well as the elastic component of the proton) is taken into account, as well as the full off-shell effects for the $\PW$ boson. Results obtained in this way are referred to as `Resonant (in)elastic' in the following Sections. \\[1mm]
\noindent \textbf{Neutrino-nucleus scattering.}
In addition to resolving the photon field of individual nucleons, the neutrino may coherently scatter upon the photon field of the entire nucleus. This can be taken into account following the techniques of~\cite{Ballett:2018uuc}, which have been applied to \PW-boson production in~\cite{Zhou:2019vxt,Beacom:2019pzs}.
The same formalism has been adopted for \PW-boson production here, using the definition of kinematics as in~\cite{Ballett:2018uuc}. This contribution will be shown for the total inclusive cross-section in the following Section, and referred to as `Resonant coherent' to distinguish it from the process discussed above.

\noindent \textbf{(Anti)neutrino-electron scattering.} 
The neutrino-nucleon scattering processes provide the dominant contributions to the cross-section for most neutrino energies. An exception occurs for the scattering of electron antineutrinos upon atomic electrons, where the process $\bar{\nu}_e e^-\to f\bar{f}^{\prime}$ becomes resonantly enhanced, which occurs for $E_{\bar{\nu}_e} \approx 6~\PeV$.

To provide an accurate prediction of the scattering rate in this energy range, the complete QCD and Electroweak (EW) corrections have been evaluated for all processes of the form $\bar{\nu}_e e^-\to f\bar{f}^{\prime}$ (including non-factorisable corrections). 
To account for the finite width effects of the $\PW$-boson at $\mathcal{O}(\alpha)$, the calculation is performed in the Complex-Mass-Scheme~\cite{Denner:2005fg} (CMS). The masses of fermions are neglected (wherever possible) throughout the calculation, with the exception that $b$- and $t$-quark masses are retained throughout. The results are obtained with the aid of~{\tt FeynArts}~\cite{Hahn:2000kx} and~{\tt FormCalc}~\cite{Hahn:1998yk} and presented in terms of complex scalar one-loop integrals. To provide differential cross-section predictions, the calculation has been implemented in a flexible~{\tt FORTRAN} code, where one-loop integrals are evaluated numerically with~{\tt OneLOop}~\cite{vanHameren:2009dr,vanHameren:2010cp}. The integration over the relevant two- and three-body phase spaces for virtual and real corrections is performed numerically with the~{\tt VEGAS} algorithm implemented in~{\tt CUBA}~\cite{Hahn:2004fe}, and the technique of Dipole-Subtraction is used to regularise the implicit and explicit infrared divergences present in the differential calculation~\cite{Catani:1996vz} (see also~\cite{Dittmaier:1999mb} for massless QED calculations).
Further refinements to the calculation are also made by including the effect of higher-order initial state radiation (ISR) corrections to the incoming electron state. These corrections are included by applying the structure function approach~\cite{Beenakker:1996kt}---note that  a recent re-calculation of this process~\cite{Blumlein:2019srk} (validating the calculation~\cite{Blumlein:2011mi}) has highlighted corrections and terms missing in the original work.
For the numerical results shown in this work, the leading logarithmic (LL) corrections up to $\mathcal{O}(\alpha^3)$  as well as the impact of soft exponentiation~\cite{Yennie:1961ad} are included~\cite{Kuraev:1985hb,Nicrosini:1986sm,Nicrosini:1987sw,Berends:1987ab} (see for example Eqns.~(5.2)-(5.6) of Ref.~\cite{Denner:2000bj}) --- results including these corrections will contain the label `+LL'.

\section{Numerical results}
In this Section, numerical predictions are provided for a few specific scattering processes encountered in the collision of (anti)neutrinos with an ${\rm H}_{2}{\rm O}$ molecule. 
These results are obtained with the following numerical inputs: $\alpha_0 = 1/137$, $\alpha_s = 0.118$, $M_W^{\rm os} = 80.385~\GeV$, $\Gamma_W^{\rm os} = 2.085~\GeV$, $M_Z^{\rm os} = 91.1876~\GeV$, $\Gamma_Z^{\rm os} = 2.4952~\GeV$, $m_b = 4.5~\GeV$, $m_t = 173~\GeV$, $m_h = 125.0~\GeV$, $m_e = 0.511~\MeV$, $G_{F} = 1.16638\cdot10^{-5}~\GeV^{-2}$. The following fermion masses are also used in the evaluation vacuum polarisation $\Delta \alpha$ (evaluated perturbatively at one-loop): $m_d = m_u = 50~\MeV$, $m_s = 150~\MeV$, $m_c = 1.5~\GeV$, $m_\mu = 105~\MeV$, $m_{\tau} = 1.78~\GeV$.
The calculation is performed in the CMS where a complex value for the weak mixing angle $\sw$ is derived according to the relation $\sw^2 = 1 - \mu_W/\mu_Z$, with $\mu_V = m_V^2 - i \Gamma_V M_V$. The $\alpha_{G_F}$-scheme is used as default throughout the calculation, where this value is derived using the above inputs while computing $\Delta r$ at one-loop. \\[1mm]
\noindent \textbf{Total inclusive cross-section.}
This study focusses on the total (summed over all final states) inclusive cross-section obtained for antineutrino collisions with either a nucleus ($\bar{\nu}+N$) or electron ($\bar{\nu}+e^-$) target of an ${\rm H}_{2}{\rm O}$ molecule. 
The predictions are produced specifically for electron antineutrino projectiles within the energy range of $E_{\bar\nu_e} \in [0.01,50]~\PeV$. The total cross-section in $\nu+N$ collisions has also been produced, and is available upon request. The cross-sections obtained for either neutrino or muon/tau antineutrinos incident upon an electron target are numerically unimportant in this energy range.

\begin{figure}[h]
  \begin{center}
    \makebox{\includegraphics[width=1.0\columnwidth]{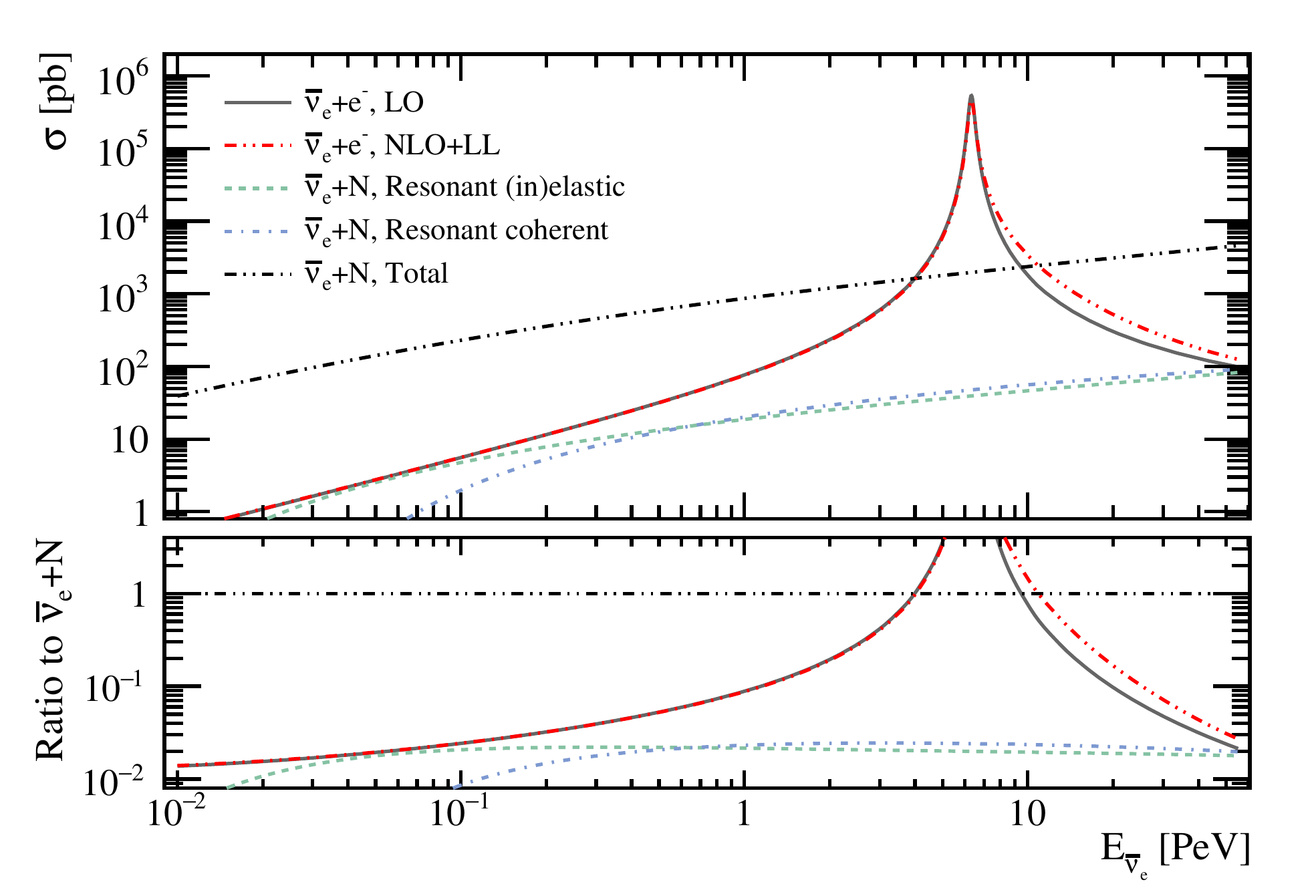}}
  \end{center}
  \vspace{-0.8cm}
  \caption{The total electron antineutrino cross-section on atomic electrons and a ${\rm H}_2{\rm O}$-nucleon as a function of incident antineutrino energy. In the case of $\bar{\nu}_e+N$ collisions, the resonant contribution generated within the photon field of the nucleus is shown separately.}
  \label{fig:Incl}
\end{figure}

The results of this study are shown in Fig.~\ref{fig:Incl}. In the upper plot, distributions for the total $\bar{\nu}_e+e^-$ cross-section (summed over all final states), the total $\bar{\nu}_e+N$ cross-section (the sum of CC, NC, and all resonant contributions), as well as the individual resonant contributions to the $\bar{\nu}_e+N$ cross-section are shown. In the lower plot, each of these distributions are shown normalised with respect to total $\bar{\nu}_e+N$ cross-section. The theoretical uncertainties of these distributions have been omitted from this Figure for visibility, but are discussed in what follows.
It should be noted that the atomic-level cross-section for ${\rm H}_{2}{\rm O}$ is obtained by multiplying $\bar{\nu}_e+N$ and $\bar{\nu}_e+e^-$ cross-sections by a factor of 18 and 10 respectively, corresponding to the mass and atomic number of the molecules constituents.

As shown in Fig.~\ref{fig:Incl}, all contributions are necessary to provide a precise (\%-accurate) prediction of the total cross-section. For multi-$\TeV$ energies and above, the (in)elastic resonant contributions amount to $2\%$, whereas the coherent contribution becomes relevant in the \PeV range and also amounts to $2\%$ of the total rate.
Within the energy range of $E_{\bar{\nu}_e} \in[4,10]~\PeV$, the $\bar{\nu}_e+e^-$ contribution dominates the total cross-section. It should also be noted that this process receives large corrections ($\approx$ factor of two) for $\sqrt{2 m_e E_{\bar{\nu}_e}} \gtrsim m_W$. The impact of radiative corrections to this process within the resonance region will be considered towards the end of this Section.

\begin{figure}[ht!]
  \begin{center}
    \makebox{\includegraphics[width=1.0\columnwidth]{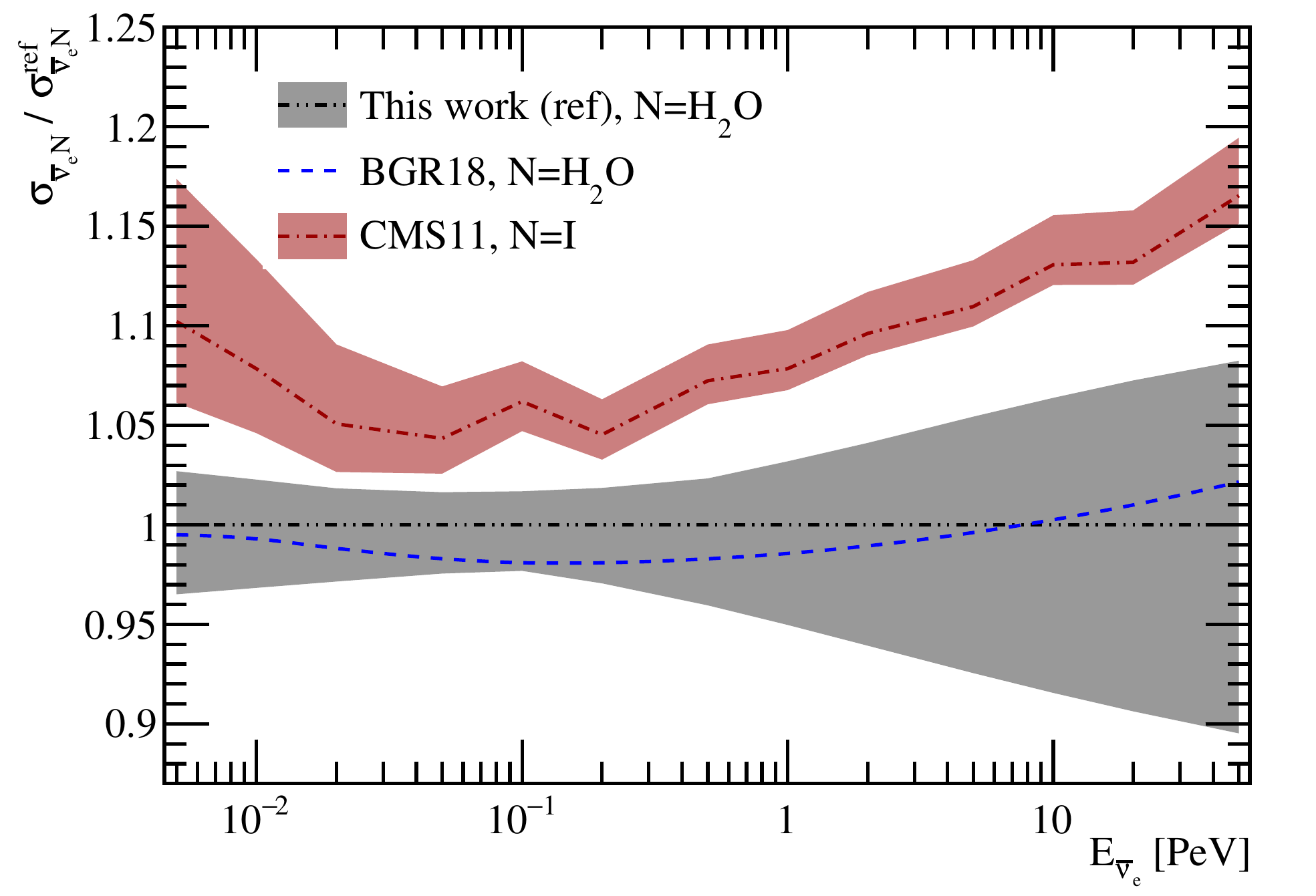}}
  \end{center}
  \vspace{-0.8cm}
  \caption{Comparison of the predictions for the inclusive cross-section in $\bar{\nu}_{e}+N$ collisions of this work with that of BGR18~\cite{Bertone:2018dse} and CMS11~\cite{CooperSarkar:2011pa}. The CMS11 results are obtained with a free isoscalar-nucleon target, while the results of this work and BGR18 include nuclear corrections.}
  \label{fig:comparison}
\end{figure}

As an additional check of the results in Fig.~\ref{fig:Incl}, the neutrino-nucleon cross-sections obtained in this work are compared to the ``BGR18"~\cite{Bertone:2018dse} and ``CMS11"~\cite{CooperSarkar:2011pa} predictions in Fig.~\ref{fig:comparison}---a comparison to other benchmark calculations~\cite{Gandhi:1998ri,Connolly:2011vc,Albacete:2015zra} was performed within Ref.~\cite{Bertone:2018dse}.
The theoretical uncertainty in this work is obtained by adding in quadrature the 1$\sigma$~CL uncertainties of the free PDFs and nuclear corrections, which are then combined linearly with the uncertainty due to scale variation. The scale variation uncertainty is obtained by varying the renormalisation scale by a factor of two around the nominal scale $Q$. A complete breakdown of these uncertainties is provided as Supplementary material.

The BGR18 calculation is consistent with the result of this work. This is not surprising since the treatment of nuclear corrections and mass effects is identical, and the input PDF set of BGR18 is the NNPDF3.1sx PDF set~\cite{Ball:2017otu}, which is extracted from the same dataset as in this work. The main differences are that the BGR18 calculation includes LHCb $D$-hadron data~\cite{Aaij:2016jht,Aaij:2013mga,Aaij:2015bpa}, and the impact of small-$x$ resummation~\cite{Bonvini:2016wki,Bonvini:2017ogt,Bonvini:2018iwt}, but does not account for QED effects. The inclusion of QED effects (to account for the resonant contributions) are most important in the multi-$\TeV$, and it is these effects which account for the difference between BGR18 and the results of this work. The BGR18 calculation is instead more suitable in the multi-$\PeV$ range ($E_{\nu} \gtrsim 50~\PeV$) where a careful description of small-$x$ dynamics is relevant.
In comparison to CMS11, the main differences are that (negative) nuclear corrections are absent and that top-quark production is treated differently. As discussed in~\cite{Bertone:2018dse}, including mass effects in top-quark production up to NLO (as in this work) leads to a large negative correction as compared to the massless calculation. Note that this comparison focuses only on the neutrino-nucleon cross-section, and the impact of coherent scattering (which can simply be added to each prediction) is not included.\\[1mm]

\begin{figure}[h]
  \begin{center}
    \makebox{\includegraphics[width=1.0\columnwidth]{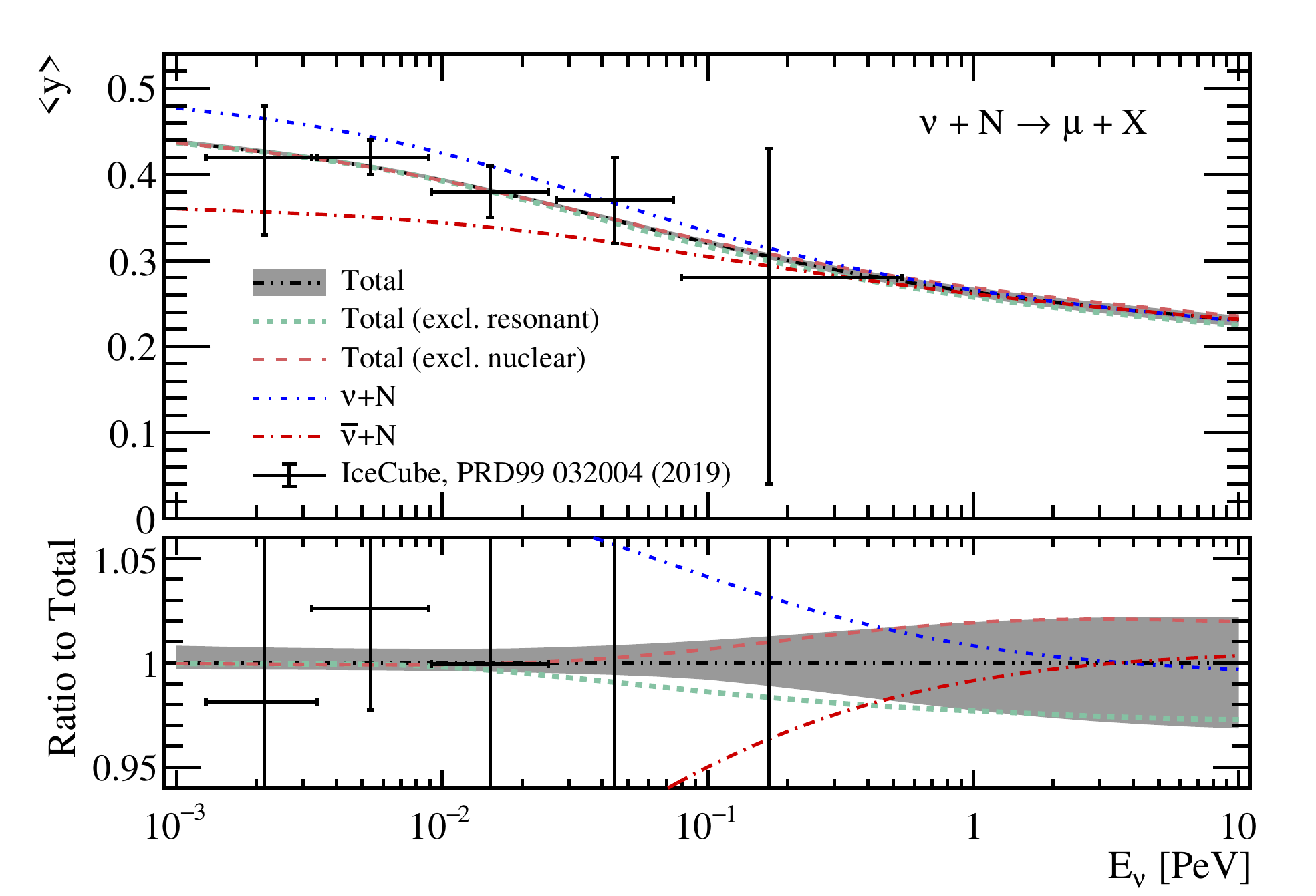}}
  \end{center}
  \vspace{-0.8cm}
  \caption{The mean inelasticity in muon production, as a function of the incident (anti)neutrino energy. In addition to the total prediction, the central values obtained when either resonant contributions or nuclear effects are excluded are also shown. For reference, the distributions obtained in either $\nu+N$ or $\bar{\nu}+N$ collisions are also shown.}
  \label{fig:ymean}
\end{figure}

\noindent \textbf{Inelasticity distribution.}
While the study of the total inclusive cross-section is theoretically instructive, it is experimentally more relevant to study exclusive final states. As a first example, the production of a charged muon in neutrino-nucleon collisions is considered (not including the coherent process). This process leads to an experimental signature composed of a cascade of hadrons and a muon, an event topology which is often referred to as a starting track.
The IceCube Collaboration has recently performed a measurement of the mean inelasticity in events of this type~\cite{Aartsen:2018vez}, which is a measure of the fractional energy transfer of the incident neutrino to hadrons. The inelasticity is equivalent to $y = 1 - E_{\mu}/E_{\nu}$ for the CC DIS process. A comparison to the available data for $\langle y \rangle$ is presented in Fig.~\ref{fig:ymean}.

The total prediction is obtained by including contributions from both $\mu^+$ and $\mu^-$ production to provide a consistent comparison with the experimental measurement~\cite{Aartsen:2018vez}. It is found that scale variation and nuclear corrections are the dominant sources of uncertainty for this distribution. The central values obtained when either the nuclear corrections or the (in)elastic resonant contributions are excluded are also shown. 

The nuclear corrections lead to a suppression of $\langle y \rangle$ by up to (2-3)\% within the $\PeV$ range. This is primarily due to the shadowing of the gluon PDF at small-$x$ values, which can lead to a suppression of CC and NC cross-sections by $\approx5\%$~\cite{Bertone:2018dse}. This suppression is largest at small-$x$ values, which at fixed-$Q^2$ corresponds to large-$y$, and thus leads to a reduced $\langle y \rangle$. It was checked that similar behaviour is observed with nNNPDF1.0 nuclear corrections obtained in Ref.~\cite{AbdulKhalek:2019mzd}.

The resonant contributions lead to an enhancement of this distribution, which almost cancels the nuclear effect discussed above. While the contribution to the total cross-section from this process is small as compared to the CC DIS contribution (below 1\%), there is still an impact on the $\langle y \rangle$ distribution due to the large inelasticity of these type of events. 
As an additional note, the prediction in Fig.~\ref{fig:ymean} is obtained using the relation $y = 1 - E_{\mu}/E_{\nu}$ which does not strictly correspond to the fractional energy transfer to hadrons for the resonant contributions, as missing energy is transferred to an outgoing neutrino. This (small) effect is not accounted for in the measurement, and is also neglected here.

It is worth mentioning that secondary lepton production (e.g. via heavy-quark production~\cite{Aartsen:2018vez,Barge:2016uzn}) may also lead to an apparent starting track type event, and subsequently alter the observed inelasticity distribution. The impact of these types of contributions is best assessed by the experimental collaborations where the impact of detector response is included.\\[1mm]
\noindent \textbf{Inclusive quark cross-section.}
The most promising channel for observing resonantly enhanced events in $\bar{\nu}_e + e^{-}$ collisions is the quark final-state, which leads to an experimental signature of hadronic showers. If the corresponding scattering process occurs within the detector, and the resultant hadronic showers are contained, a resonant peak can be constructed in the visible energy spectrum~\cite{Aartsen:2013jla}. In fact, a candidate event of this type has been observed in a data sample of partially contained hadronic showers~\cite{Lu:2017nti}.

Assuming a combined (and equal) flux of electron neutrino and antineutrinos of $E^2 \Phi_{\nu_e+\bar{\nu}_e} = 1\times10^{-8}\GeV\cdot {\rm cm}^{-2}\cdot{\rm s}^{-1}\cdot{\rm sr}^{-1}$, the IceCube Collaboration has previously estimated that $0.9$ contained hadronic shower events may be detected per year~\cite{Aartsen:2013jla} with the current detector. A more conservative estimate of $0.35$ events per year was given in Ref.~\cite{Biehl:2016psj}. 
The event rate prediction for this process is proportional to the effective detector volume, electron antineutrino flux, cross-section, as well as the time exposure. A measurement of this process is therefore most feasible with IceCube-Gen2, where an increase in the effective detector volume by a factor of 5 to 12 would be realised~\cite{Aartsen:2014njl}. It is also clear that the eventual interpretation of such a measurement will depend on a reliable prediction for the cross-section.

\begin{figure}[h]
  \begin{center}
    \makebox{\includegraphics[width=1.0\columnwidth]{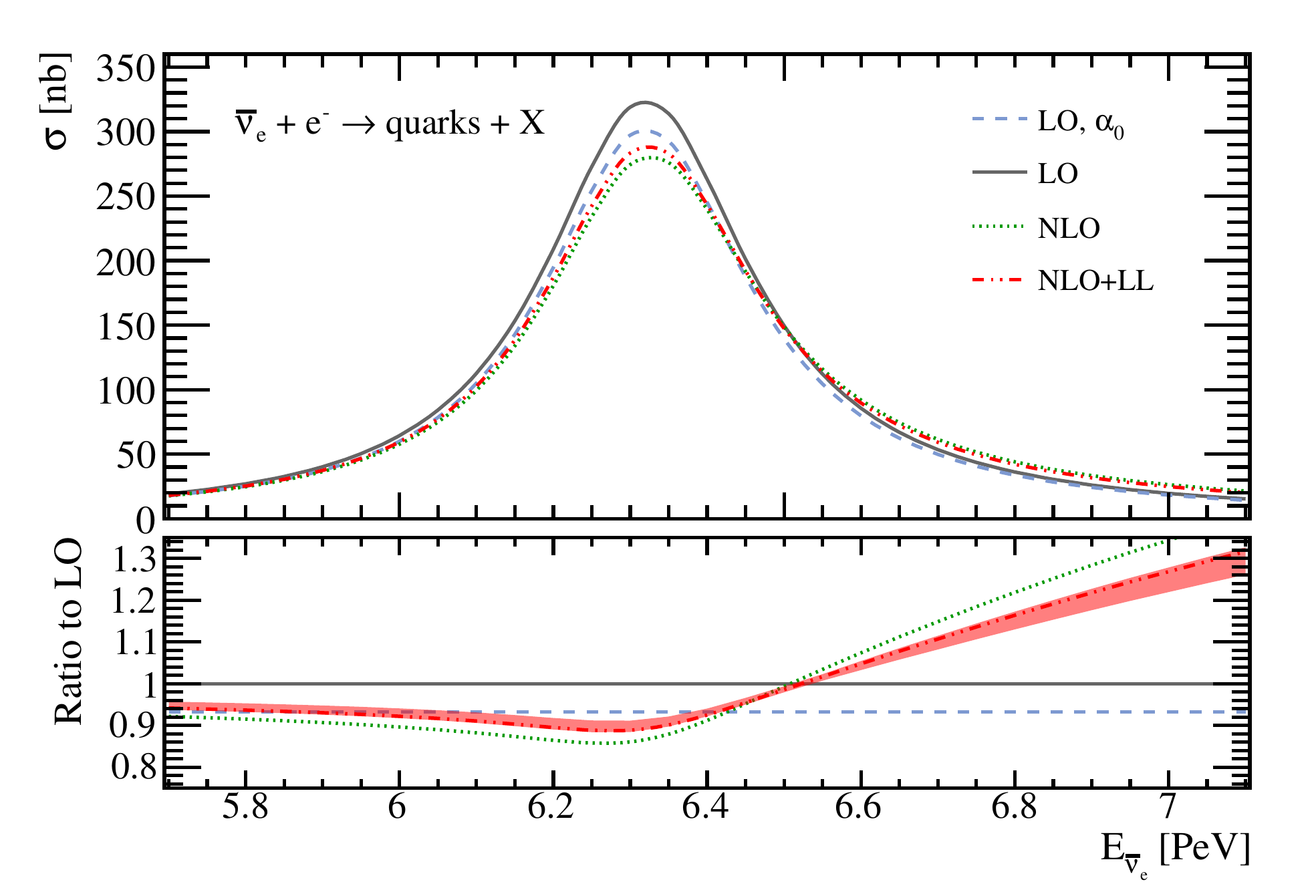}}
  \end{center}
  \vspace{-0.8cm}
  \caption{The inclusive cross-section for quark production in $\bar{\nu}_{e}+e^{-}$ collisions as a function of the incident antineutrino energy.}
  \label{fig:quarks}
\end{figure}

Such predictions, for the inclusive cross-section, are presented in Fig.~\ref{fig:quarks} as a function of the incident antineutrino energy. The central values obtained in the $\alpha_{G_F}$-scheme at various accuracies are shown, and in the lower plot these predictions are normalised by the LO prediction---obtained when the cross-section is computed in terms of $G_{F}$~\cite{Mikaelian:1980vd} via the replacement $\pi \alpha/\sw^2 \to \sqrt{2} G_F \mu_W$. For reference, the LO prediction computed in the $\alpha_0$-scheme is also shown. In the lower plot, the uncertainty of the ${\rm NLO+LL}$ prediction is shown, which is computed as the envelope of the uncertainty due scale variation (which is applied to predictions obtained when the QCD corrections are included in either a multiplicative or additive to the EW-corrected cross-section) as well as an uncertainty due input scheme choice. The uncertainty due to factorisation scale dependence (which results from including higher-order ISR corrections via structure functions) is also assessed when changing the reference scale from $\mu = \sqrt{S}$ and $\mu = \sqrt{-t}$, where $t = (p_e - p_{\bar{q}})^2$. The uncertainty due to the choice of scheme is evaluated at NLO as $\delta^{\rm scheme} = \sigma^{\rm NLO,(\alpha_0)}-\sigma^{\rm NLO}$.
The central value of the ${\rm NLO+LL}$ prediction is obtained with $\mu_F = \sqrt{-t}$ and the multiplicative prescription for the QCD corrections. In addition to the distributions in Fig.~\ref{fig:quarks}, the inclusive cross-section in bins of $E_{\bar{\nu}_e}$ is also provided in Table~\ref{tab:quarks}.

\renewcommand*{\arraystretch}{1.2}
\begin{table}[h!]
	\centering
	\begin{tabular}{c | c | c | c | c@{}}
      	\hline
	$E_{\bar{\nu}_e} [\PeV]$ & ${\rm LO(\alpha_0)}$  & ${\rm LO}$ & ${\rm NLO}$ & ${\rm NLO+LL}$\\
	\hline
$[5.8,6.0]$ & $39.16$ & $ 42.02$ & $38.04$ & $38.82^{+0.7}_{-0.2}$ \\
$[6.0,6.2]$ & $112.5$ & $ 120.8$ & $106.1$ & $108.7^{+2.3}_{-0.6}$ \\
$[6.2,6.4]$ & $268.4$ & $ 288.0$ & $250.7$ & $255.8^{+5.6}_{-1.2}$ \\
$[6.4,6.6]$ & $147.6$ & $ 158.4$ & $154.9$ & $153.8^{+1.2}_{-0.9}$ \\
$[6.6,6.8]$ & $52.10$ & $ 55.91$ & $63.61$ & $61.76^{+0.4}_{-1.4}$ \\
        \hline
	\end{tabular}
        \caption{The inclusive cross-section (in $nb$) for quark production in $\bar{\nu}_e +e^-$ scattering for several bins of  $E_{\bar{\nu}_e}$. The predictions are obtained in the $\alpha_{G_F}$-scheme, except those labelled with ``$\alpha_0$" as described in the text.}\label{tab:quarks}
\end{table}

The higher-order corrections have an important impact on the line-shape around the resonance region. In the peak region, the ${\rm NLO+LL}$ prediction receives a correction of $\approx -10\%$ as compared to ${\rm LO}$. 
The impact of the corrections on the expected event rate (which is more experimentally relevant) can also be approximated by integrating the cross-section within the resonance region, under the assumption $\Phi_{\bar{\nu}_e}\propto E^{-2}$. Using the cross-section values given in Table~\ref{tab:quarks} gives
\begin{equation}
\frac{N^{\rm NLO+LL}}{N^{{\rm LO~}}}-1.0 = -0.07\,, \quad E_{\bar{\nu}_e}  \in [5.8,6.8]~\PeV\,.
\end{equation}
While naively using $G_F$ as an input for the LO cross-section is sensible (it naturally includes universal higher-order EW effects), such an approach provides a poor approximation to the total cross-section as the dominant ISR corrections are absent. These corrections receive a logarithmic enhancement of the form $\alpha/\pi \ln\left[m_W/m_e\right]$ (which multiplies the splitting function $P_{ee}$), and amount to $\approx -(10-15)\%$ of the LO cross-section. The uncertainties of the ${\rm NLO+LL}$ predictions are sufficiently small ($\approx 1\%$) that they can be ignored as compared to the expected precision of the data.

\section{Discussion and conclusions}
In this work, predictions for multi-$\TeV$- and $\PeV$-energy neutrino scattering rates on both nuclear and electron targets has been revisited.

For neutrino-nucleus scattering, these predictions have been improved by including the resonant contributions which are generated within the photon field of the nucleus. These contributions have been included by generating leptons PDFs within bound nucleons, using a PDF set based on the luxQED formalism as a boundary condition, and are found to impact the considered distributions by $\approx (2-3)\%$. 
With respect to other determinations~\cite{Alikhanov:2015kla,Zhou:2019vxt,Beacom:2019pzs}, these predictions are made available for exclusive final states ($f\bar{f}^{\prime}$) differential in the outgoing fermion kinematics (and therefore fully account for off-shell effects), include the dominant corrections at $\mathcal{O}(\alpha)$, and additionally include a resummation of the leading collinear logarithms.
The coherent scattering process (for an Oxygen nucleus) has been included for \PW-boson production following~\cite{Ballett:2018uuc,Zhou:2019vxt,Beacom:2019pzs}, which additionally increases the total cross-section by 2\% for $\PeV$-energy (anti)electron neutrinos.

A precise calculation of $\bar{\nu}_e+e^{-}\to f\bar{f}^{\prime}$ scattering rates has also been presented, which is {\rm NLO}-accurate and additionally includes the impact of universal higher-order ISR corrections. For quark final states, the inclusive cross-section receives a correction of $\approx-10\%$ in the resonance region and will be relevant for the interpretation of $\PeV$-energy event rates at Neutrino Telescopes.

While only a select number of (mostly inclusive) observables have been presented in this work, the calculations have been implemented in such a way that fully differential predictions can be produced. It is anticipated that these calculations can also be useful as a tool for the experimental collaborations. 
In particular, interfacing~\cite{Frixione:2002ik,Nason:2004rx,Frixione:2007vw} the fixed-order calculations presented in this work with a fully exclusive Parton Shower would allow the experimental collaborations to have a more accurate modelling of both QCD and QED radiation in event simulations, which may lead to improved sensitivity of the experimental measurements. This is foreseen for future work.

{\it Acknowledgements.}
I am grateful to Stefan Dittmaier for  providing useful comments regarding the implementation of ISR QED corrections, and to Alfonso Garcia for discussions about several experimental aspects of this work. I also acknowledge previous collaboration with Valerio Bertone and Juan Rojo which was relevant to this work. This work is supported by the Dutch Organization for Scientific Research (NWO) through the VENI grant 680-47-461.

\bibliography{UHEsig}

\begin{thebibliography}{65}
\expandafter\ifx\csname natexlab\endcsname\relax\def\natexlab#1{#1}\fi
\expandafter\ifx\csname bibnamefont\endcsname\relax
  \def\bibnamefont#1{#1}\fi
\expandafter\ifx\csname bibfnamefont\endcsname\relax
  \def\bibfnamefont#1{#1}\fi
\expandafter\ifx\csname citenamefont\endcsname\relax
  \def\citenamefont#1{#1}\fi
\expandafter\ifx\csname url\endcsname\relax
  \def\url#1{\texttt{#1}}\fi
\expandafter\ifx\csname urlprefix\endcsname\relax\def\urlprefix{URL }\fi
\providecommand{\bibinfo}[2]{#2}
\providecommand{\eprint}[2][]{\url{#2}}

\bibitem[{Ice(2018)}]{IceCube:2018cha}
\bibinfo{journal}{Science} \textbf{\bibinfo{volume}{361}}, \bibinfo{pages}{147}
  (\bibinfo{year}{2018}).

\bibitem[{\citenamefont{Aartsen et~al.}(2018)}]{IceCube:2018dnn}
\bibinfo{author}{\bibfnamefont{M.~G.} \bibnamefont{Aartsen}}
  \bibnamefont{et~al.} (\bibinfo{collaboration}{IceCube, Fermi-LAT, MAGIC,
  AGILE, ASAS-SN, HAWC, H.E.S.S., INTEGRAL, Kanata, Kiso, Kapteyn, Liverpool
  Telescope, Subaru, Swift NuSTAR, VERITAS, VLA/17B-403}),
  \bibinfo{journal}{Science} \textbf{\bibinfo{volume}{361}},
  \bibinfo{pages}{eaat1378} (\bibinfo{year}{2018}), \eprint{1807.08816}.

\bibitem[{\citenamefont{Aartsen et~al.}(2017{\natexlab{a}})}]{Aartsen:2017nbu}
\bibinfo{author}{\bibfnamefont{M.~G.} \bibnamefont{Aartsen}}
  \bibnamefont{et~al.} (\bibinfo{collaboration}{IceCube}),
  \bibinfo{journal}{Eur. Phys. J.} \textbf{\bibinfo{volume}{C77}},
  \bibinfo{pages}{692} (\bibinfo{year}{2017}{\natexlab{a}}),
  \eprint{1705.07780}.

\bibitem[{\citenamefont{Aartsen et~al.}(2017{\natexlab{b}})}]{Aartsen:2017kpd}
\bibinfo{author}{\bibfnamefont{M.~G.} \bibnamefont{Aartsen}}
  \bibnamefont{et~al.} (\bibinfo{collaboration}{IceCube}),
  \bibinfo{journal}{Nature} \textbf{\bibinfo{volume}{551}},
  \bibinfo{pages}{596} (\bibinfo{year}{2017}{\natexlab{b}}),
  \eprint{1711.08119}.

\bibitem[{\citenamefont{Bustamante and Connolly}(2019)}]{Bustamante:2017xuy}
\bibinfo{author}{\bibfnamefont{M.}~\bibnamefont{Bustamante}} \bibnamefont{and}
  \bibinfo{author}{\bibfnamefont{A.}~\bibnamefont{Connolly}},
  \bibinfo{journal}{Phys. Rev. Lett.} \textbf{\bibinfo{volume}{122}},
  \bibinfo{pages}{041101} (\bibinfo{year}{2019}), \eprint{1711.11043}.

\bibitem[{\citenamefont{Aartsen et~al.}(2019)}]{Aartsen:2018vez}
\bibinfo{author}{\bibfnamefont{M.~G.} \bibnamefont{Aartsen}}
  \bibnamefont{et~al.} (\bibinfo{collaboration}{IceCube}),
  \bibinfo{journal}{Phys. Rev.} \textbf{\bibinfo{volume}{D99}},
  \bibinfo{pages}{032004} (\bibinfo{year}{2019}), \eprint{1808.07629}.

\bibitem[{\citenamefont{Glashow}(1960)}]{Glashow:1960zz}
\bibinfo{author}{\bibfnamefont{S.~L.} \bibnamefont{Glashow}},
  \bibinfo{journal}{Phys. Rev.} \textbf{\bibinfo{volume}{118}},
  \bibinfo{pages}{316} (\bibinfo{year}{1960}).

\bibitem[{\citenamefont{Anchordoqui et~al.}(2005)\citenamefont{Anchordoqui,
  Goldberg, Halzen, and Weiler}}]{Anchordoqui:2004eb}
\bibinfo{author}{\bibfnamefont{L.~A.} \bibnamefont{Anchordoqui}},
  \bibinfo{author}{\bibfnamefont{H.}~\bibnamefont{Goldberg}},
  \bibinfo{author}{\bibfnamefont{F.}~\bibnamefont{Halzen}}, \bibnamefont{and}
  \bibinfo{author}{\bibfnamefont{T.~J.} \bibnamefont{Weiler}},
  \bibinfo{journal}{Phys. Lett.} \textbf{\bibinfo{volume}{B621}},
  \bibinfo{pages}{18} (\bibinfo{year}{2005}), \eprint{hep-ph/0410003}.

\bibitem[{\citenamefont{Hummer et~al.}(2010)\citenamefont{Hummer, Maltoni,
  Winter, and Yaguna}}]{Hummer:2010ai}
\bibinfo{author}{\bibfnamefont{S.}~\bibnamefont{Hummer}},
  \bibinfo{author}{\bibfnamefont{M.}~\bibnamefont{Maltoni}},
  \bibinfo{author}{\bibfnamefont{W.}~\bibnamefont{Winter}}, \bibnamefont{and}
  \bibinfo{author}{\bibfnamefont{C.}~\bibnamefont{Yaguna}},
  \bibinfo{journal}{Astropart. Phys.} \textbf{\bibinfo{volume}{34}},
  \bibinfo{pages}{205} (\bibinfo{year}{2010}), \eprint{1007.0006}.

\bibitem[{\citenamefont{Xing and Zhou}(2011)}]{Xing:2011zm}
\bibinfo{author}{\bibfnamefont{Z.-z.} \bibnamefont{Xing}} \bibnamefont{and}
  \bibinfo{author}{\bibfnamefont{S.}~\bibnamefont{Zhou}},
  \bibinfo{journal}{Phys. Rev.} \textbf{\bibinfo{volume}{D84}},
  \bibinfo{pages}{033006} (\bibinfo{year}{2011}), \eprint{1105.4114}.

\bibitem[{\citenamefont{Biehl et~al.}(2017)\citenamefont{Biehl, Fedynitch,
  Palladino, Weiler, and Winter}}]{Biehl:2016psj}
\bibinfo{author}{\bibfnamefont{D.}~\bibnamefont{Biehl}},
  \bibinfo{author}{\bibfnamefont{A.}~\bibnamefont{Fedynitch}},
  \bibinfo{author}{\bibfnamefont{A.}~\bibnamefont{Palladino}},
  \bibinfo{author}{\bibfnamefont{T.~J.} \bibnamefont{Weiler}},
  \bibnamefont{and} \bibinfo{author}{\bibfnamefont{W.}~\bibnamefont{Winter}},
  \bibinfo{journal}{JCAP} \textbf{\bibinfo{volume}{1701}}, \bibinfo{pages}{033}
  (\bibinfo{year}{2017}), \eprint{1611.07983}.

\bibitem[{\citenamefont{Adrian-Martinez
  et~al.}(2016)}]{Adrian-Martinez:2016fdl}
\bibinfo{author}{\bibfnamefont{S.}~\bibnamefont{Adrian-Martinez}}
  \bibnamefont{et~al.} (\bibinfo{collaboration}{KM3Net}), \bibinfo{journal}{J.
  Phys.} \textbf{\bibinfo{volume}{G43}}, \bibinfo{pages}{084001}
  (\bibinfo{year}{2016}), \eprint{1601.07459}.

\bibitem[{\citenamefont{Avrorin et~al.}(2018)}]{Avrorin:2018ijk}
\bibinfo{author}{\bibfnamefont{A.~D.} \bibnamefont{Avrorin}}
  \bibnamefont{et~al.} (\bibinfo{collaboration}{Baikal-GVD}),
  \bibinfo{journal}{EPJ Web Conf.} \textbf{\bibinfo{volume}{191}},
  \bibinfo{pages}{01006} (\bibinfo{year}{2018}), \eprint{1808.10353}.

\bibitem[{\citenamefont{Bertone et~al.}(2019)\citenamefont{Bertone, Gauld, and
  Rojo}}]{Bertone:2018dse}
\bibinfo{author}{\bibfnamefont{V.}~\bibnamefont{Bertone}},
  \bibinfo{author}{\bibfnamefont{R.}~\bibnamefont{Gauld}}, \bibnamefont{and}
  \bibinfo{author}{\bibfnamefont{J.}~\bibnamefont{Rojo}},
  \bibinfo{journal}{JHEP} \textbf{\bibinfo{volume}{01}}, \bibinfo{pages}{217}
  (\bibinfo{year}{2019}), \eprint{1808.02034}.

\bibitem[{\citenamefont{Seckel}(1998)}]{Seckel:1997kk}
\bibinfo{author}{\bibfnamefont{D.}~\bibnamefont{Seckel}},
  \bibinfo{journal}{Phys. Rev. Lett.} \textbf{\bibinfo{volume}{80}},
  \bibinfo{pages}{900} (\bibinfo{year}{1998}), \eprint{hep-ph/9709290}.

\bibitem[{\citenamefont{Alikhanov}(2016)}]{Alikhanov:2015kla}
\bibinfo{author}{\bibfnamefont{I.}~\bibnamefont{Alikhanov}},
  \bibinfo{journal}{Phys. Lett.} \textbf{\bibinfo{volume}{B756}},
  \bibinfo{pages}{247} (\bibinfo{year}{2016}), \eprint{1503.08817}.

\bibitem[{\citenamefont{Bertone et~al.}(2018)\citenamefont{Bertone, Carrazza,
  Hartland, and Rojo}}]{Bertone:2017bme}
\bibinfo{author}{\bibfnamefont{V.}~\bibnamefont{Bertone}},
  \bibinfo{author}{\bibfnamefont{S.}~\bibnamefont{Carrazza}},
  \bibinfo{author}{\bibfnamefont{N.~P.} \bibnamefont{Hartland}},
  \bibnamefont{and} \bibinfo{author}{\bibfnamefont{J.}~\bibnamefont{Rojo}}
  (\bibinfo{collaboration}{NNPDF}), \bibinfo{journal}{SciPost Phys.}
  \textbf{\bibinfo{volume}{5}}, \bibinfo{pages}{008} (\bibinfo{year}{2018}),
  \eprint{1712.07053}.

\bibitem[{\citenamefont{Manohar et~al.}(2016)\citenamefont{Manohar, Nason,
  Salam, and Zanderighi}}]{Manohar:2016nzj}
\bibinfo{author}{\bibfnamefont{A.}~\bibnamefont{Manohar}},
  \bibinfo{author}{\bibfnamefont{P.}~\bibnamefont{Nason}},
  \bibinfo{author}{\bibfnamefont{G.~P.} \bibnamefont{Salam}}, \bibnamefont{and}
  \bibinfo{author}{\bibfnamefont{G.}~\bibnamefont{Zanderighi}},
  \bibinfo{journal}{Phys. Rev. Lett.} \textbf{\bibinfo{volume}{117}},
  \bibinfo{pages}{242002} (\bibinfo{year}{2016}), \eprint{1607.04266}.

\bibitem[{\citenamefont{Manohar et~al.}(2017)\citenamefont{Manohar, Nason,
  Salam, and Zanderighi}}]{Manohar:2017eqh}
\bibinfo{author}{\bibfnamefont{A.~V.} \bibnamefont{Manohar}},
  \bibinfo{author}{\bibfnamefont{P.}~\bibnamefont{Nason}},
  \bibinfo{author}{\bibfnamefont{G.~P.} \bibnamefont{Salam}}, \bibnamefont{and}
  \bibinfo{author}{\bibfnamefont{G.}~\bibnamefont{Zanderighi}},
  \bibinfo{journal}{JHEP} \textbf{\bibinfo{volume}{12}}, \bibinfo{pages}{046}
  (\bibinfo{year}{2017}), \eprint{1708.01256}.

\bibitem[{\citenamefont{Bertone et~al.}(2014)\citenamefont{Bertone, Carrazza,
  and Rojo}}]{Bertone:2013vaa}
\bibinfo{author}{\bibfnamefont{V.}~\bibnamefont{Bertone}},
  \bibinfo{author}{\bibfnamefont{S.}~\bibnamefont{Carrazza}}, \bibnamefont{and}
  \bibinfo{author}{\bibfnamefont{J.}~\bibnamefont{Rojo}},
  \bibinfo{journal}{Comput.Phys.Commun.} \textbf{\bibinfo{volume}{185}},
  \bibinfo{pages}{1647} (\bibinfo{year}{2014}), \eprint{1310.1394}.

\bibitem[{\citenamefont{Bertone et~al.}(2015)\citenamefont{Bertone, Carrazza,
  Pagani, and Zaro}}]{Bertone:2015lqa}
\bibinfo{author}{\bibfnamefont{V.}~\bibnamefont{Bertone}},
  \bibinfo{author}{\bibfnamefont{S.}~\bibnamefont{Carrazza}},
  \bibinfo{author}{\bibfnamefont{D.}~\bibnamefont{Pagani}}, \bibnamefont{and}
  \bibinfo{author}{\bibfnamefont{M.}~\bibnamefont{Zaro}},
  \bibinfo{journal}{JHEP} \textbf{\bibinfo{volume}{11}}, \bibinfo{pages}{194}
  (\bibinfo{year}{2015}), \eprint{1508.07002}.

\bibitem[{\citenamefont{Forte et~al.}(2010)\citenamefont{Forte, Laenen, Nason,
  and Rojo}}]{Forte:2010ta}
\bibinfo{author}{\bibfnamefont{S.}~\bibnamefont{Forte}},
  \bibinfo{author}{\bibfnamefont{E.}~\bibnamefont{Laenen}},
  \bibinfo{author}{\bibfnamefont{P.}~\bibnamefont{Nason}}, \bibnamefont{and}
  \bibinfo{author}{\bibfnamefont{J.}~\bibnamefont{Rojo}},
  \bibinfo{journal}{Nucl. Phys.} \textbf{\bibinfo{volume}{B834}},
  \bibinfo{pages}{116} (\bibinfo{year}{2010}), \eprint{1001.2312}.

\bibitem[{\citenamefont{Ball et~al.}(2016)\citenamefont{Ball, Bertone, Bonvini,
  Forte, Groth~Merrild, Rojo, and Rottoli}}]{Ball:2015tna}
\bibinfo{author}{\bibfnamefont{R.~D.} \bibnamefont{Ball}},
  \bibinfo{author}{\bibfnamefont{V.}~\bibnamefont{Bertone}},
  \bibinfo{author}{\bibfnamefont{M.}~\bibnamefont{Bonvini}},
  \bibinfo{author}{\bibfnamefont{S.}~\bibnamefont{Forte}},
  \bibinfo{author}{\bibfnamefont{P.}~\bibnamefont{Groth~Merrild}},
  \bibinfo{author}{\bibfnamefont{J.}~\bibnamefont{Rojo}}, \bibnamefont{and}
  \bibinfo{author}{\bibfnamefont{L.}~\bibnamefont{Rottoli}},
  \bibinfo{journal}{Phys. Lett.} \textbf{\bibinfo{volume}{B754}},
  \bibinfo{pages}{49} (\bibinfo{year}{2016}), \eprint{1510.00009}.

\bibitem[{\citenamefont{Eskola et~al.}(2017)\citenamefont{Eskola, Paakkinen,
  Paukkunen, and Salgado}}]{Eskola:2016oht}
\bibinfo{author}{\bibfnamefont{K.~J.} \bibnamefont{Eskola}},
  \bibinfo{author}{\bibfnamefont{P.}~\bibnamefont{Paakkinen}},
  \bibinfo{author}{\bibfnamefont{H.}~\bibnamefont{Paukkunen}},
  \bibnamefont{and} \bibinfo{author}{\bibfnamefont{C.~A.}
  \bibnamefont{Salgado}}, \bibinfo{journal}{Eur. Phys. J.}
  \textbf{\bibinfo{volume}{C77}}, \bibinfo{pages}{163} (\bibinfo{year}{2017}),
  \eprint{1612.05741}.

\bibitem[{\citenamefont{Dulat et~al.}(2016)\citenamefont{Dulat, Hou, Gao,
  Guzzi, Huston, Nadolsky, Pumplin, Schmidt, Stump, and Yuan}}]{Dulat:2015mca}
\bibinfo{author}{\bibfnamefont{S.}~\bibnamefont{Dulat}},
  \bibinfo{author}{\bibfnamefont{T.-J.} \bibnamefont{Hou}},
  \bibinfo{author}{\bibfnamefont{J.}~\bibnamefont{Gao}},
  \bibinfo{author}{\bibfnamefont{M.}~\bibnamefont{Guzzi}},
  \bibinfo{author}{\bibfnamefont{J.}~\bibnamefont{Huston}},
  \bibinfo{author}{\bibfnamefont{P.}~\bibnamefont{Nadolsky}},
  \bibinfo{author}{\bibfnamefont{J.}~\bibnamefont{Pumplin}},
  \bibinfo{author}{\bibfnamefont{C.}~\bibnamefont{Schmidt}},
  \bibinfo{author}{\bibfnamefont{D.}~\bibnamefont{Stump}}, \bibnamefont{and}
  \bibinfo{author}{\bibfnamefont{C.~P.} \bibnamefont{Yuan}},
  \bibinfo{journal}{Phys. Rev.} \textbf{\bibinfo{volume}{D93}},
  \bibinfo{pages}{033006} (\bibinfo{year}{2016}), \eprint{1506.07443}.

\bibitem[{\citenamefont{Ballett et~al.}(2019)\citenamefont{Ballett, Hostert,
  Pascoli, Perez-Gonzalez, Tabrizi, and Zukanovich~Funchal}}]{Ballett:2018uuc}
\bibinfo{author}{\bibfnamefont{P.}~\bibnamefont{Ballett}},
  \bibinfo{author}{\bibfnamefont{M.}~\bibnamefont{Hostert}},
  \bibinfo{author}{\bibfnamefont{S.}~\bibnamefont{Pascoli}},
  \bibinfo{author}{\bibfnamefont{Y.~F.} \bibnamefont{Perez-Gonzalez}},
  \bibinfo{author}{\bibfnamefont{Z.}~\bibnamefont{Tabrizi}}, \bibnamefont{and}
  \bibinfo{author}{\bibfnamefont{R.}~\bibnamefont{Zukanovich~Funchal}},
  \bibinfo{journal}{JHEP} \textbf{\bibinfo{volume}{01}}, \bibinfo{pages}{119}
  (\bibinfo{year}{2019}), \eprint{1807.10973}.

\bibitem[{\citenamefont{Zhou and Beacom}(2020{\natexlab{a}})}]{Zhou:2019vxt}
\bibinfo{author}{\bibfnamefont{B.}~\bibnamefont{Zhou}} \bibnamefont{and}
  \bibinfo{author}{\bibfnamefont{J.~F.} \bibnamefont{Beacom}},
  \bibinfo{journal}{Phys. Rev. D} \textbf{\bibinfo{volume}{101}},
  \bibinfo{pages}{036011} (\bibinfo{year}{2020}{\natexlab{a}}),
  \eprint{1910.08090}.

\bibitem[{\citenamefont{Zhou and Beacom}(2020{\natexlab{b}})}]{Beacom:2019pzs}
\bibinfo{author}{\bibfnamefont{B.}~\bibnamefont{Zhou}} \bibnamefont{and}
  \bibinfo{author}{\bibfnamefont{J.~F.} \bibnamefont{Beacom}},
  \bibinfo{journal}{Phys. Rev. D} \textbf{\bibinfo{volume}{101}},
  \bibinfo{pages}{036010} (\bibinfo{year}{2020}{\natexlab{b}}),
  \eprint{1910.10720}.

\bibitem[{\citenamefont{Denner et~al.}(2005)\citenamefont{Denner, Dittmaier,
  Roth, and Wieders}}]{Denner:2005fg}
\bibinfo{author}{\bibfnamefont{A.}~\bibnamefont{Denner}},
  \bibinfo{author}{\bibfnamefont{S.}~\bibnamefont{Dittmaier}},
  \bibinfo{author}{\bibfnamefont{M.}~\bibnamefont{Roth}}, \bibnamefont{and}
  \bibinfo{author}{\bibfnamefont{L.~H.} \bibnamefont{Wieders}},
  \bibinfo{journal}{Nucl. Phys.} \textbf{\bibinfo{volume}{B724}},
  \bibinfo{pages}{247} (\bibinfo{year}{2005}), \bibinfo{note}{[Erratum: Nucl.
  Phys.B854,504(2012)]}, \eprint{hep-ph/0505042}.

\bibitem[{\citenamefont{Hahn}(2001)}]{Hahn:2000kx}
\bibinfo{author}{\bibfnamefont{T.}~\bibnamefont{Hahn}},
  \bibinfo{journal}{Comput. Phys. Commun.} \textbf{\bibinfo{volume}{140}},
  \bibinfo{pages}{418} (\bibinfo{year}{2001}), \eprint{hep-ph/0012260}.

\bibitem[{\citenamefont{Hahn and Perez-Victoria}(1999)}]{Hahn:1998yk}
\bibinfo{author}{\bibfnamefont{T.}~\bibnamefont{Hahn}} \bibnamefont{and}
  \bibinfo{author}{\bibfnamefont{M.}~\bibnamefont{Perez-Victoria}},
  \bibinfo{journal}{Comput. Phys. Commun.} \textbf{\bibinfo{volume}{118}},
  \bibinfo{pages}{153} (\bibinfo{year}{1999}), \eprint{hep-ph/9807565}.

\bibitem[{\citenamefont{van Hameren et~al.}(2009)\citenamefont{van Hameren,
  Papadopoulos, and Pittau}}]{vanHameren:2009dr}
\bibinfo{author}{\bibfnamefont{A.}~\bibnamefont{van Hameren}},
  \bibinfo{author}{\bibfnamefont{C.~G.} \bibnamefont{Papadopoulos}},
  \bibnamefont{and} \bibinfo{author}{\bibfnamefont{R.}~\bibnamefont{Pittau}},
  \bibinfo{journal}{JHEP} \textbf{\bibinfo{volume}{09}}, \bibinfo{pages}{106}
  (\bibinfo{year}{2009}), \eprint{0903.4665}.

\bibitem[{\citenamefont{van Hameren}(2011)}]{vanHameren:2010cp}
\bibinfo{author}{\bibfnamefont{A.}~\bibnamefont{van Hameren}},
  \bibinfo{journal}{Comput. Phys. Commun.} \textbf{\bibinfo{volume}{182}},
  \bibinfo{pages}{2427} (\bibinfo{year}{2011}), \eprint{1007.4716}.

\bibitem[{\citenamefont{Hahn}(2005)}]{Hahn:2004fe}
\bibinfo{author}{\bibfnamefont{T.}~\bibnamefont{Hahn}},
  \bibinfo{journal}{Comput. Phys. Commun.} \textbf{\bibinfo{volume}{168}},
  \bibinfo{pages}{78} (\bibinfo{year}{2005}), \eprint{hep-ph/0404043}.

\bibitem[{\citenamefont{Catani and Seymour}(1997)}]{Catani:1996vz}
\bibinfo{author}{\bibfnamefont{S.}~\bibnamefont{Catani}} \bibnamefont{and}
  \bibinfo{author}{\bibfnamefont{M.~H.} \bibnamefont{Seymour}},
  \bibinfo{journal}{Nucl. Phys.} \textbf{\bibinfo{volume}{B485}},
  \bibinfo{pages}{291} (\bibinfo{year}{1997}), \bibinfo{note}{[Erratum: Nucl.
  Phys.B510,503(1998)]}, \eprint{hep-ph/9605323}.

\bibitem[{\citenamefont{Dittmaier}(2000)}]{Dittmaier:1999mb}
\bibinfo{author}{\bibfnamefont{S.}~\bibnamefont{Dittmaier}},
  \bibinfo{journal}{Nucl. Phys.} \textbf{\bibinfo{volume}{B565}},
  \bibinfo{pages}{69} (\bibinfo{year}{2000}), \eprint{hep-ph/9904440}.

\bibitem[{\citenamefont{Beenakker et~al.}(1996)}]{Beenakker:1996kt}
\bibinfo{author}{\bibfnamefont{W.}~\bibnamefont{Beenakker}}
  \bibnamefont{et~al.}, in \emph{\bibinfo{booktitle}{CERN Workshop on LEP2
  \\Physics (followed by 2nd meeting, 15-16 Jun 1995 and \\3rd meeting 2-3 Nov
  1995) Geneva, Switzerland, \\ February 2-3, 1995}} (\bibinfo{year}{1996}),
  pp. \bibinfo{pages}{79--139}, \eprint{hep-ph/9602351}.

\bibitem[{\citenamefont{Blumlein et~al.}(2019)\citenamefont{Blumlein,
  De~Freitas, Raab, and Schonwald}}]{Blumlein:2019srk}
\bibinfo{author}{\bibfnamefont{J.}~\bibnamefont{Blumlein}},
  \bibinfo{author}{\bibfnamefont{A.}~\bibnamefont{De~Freitas}},
  \bibinfo{author}{\bibfnamefont{C.~G.} \bibnamefont{Raab}}, \bibnamefont{and}
  \bibinfo{author}{\bibfnamefont{K.}~\bibnamefont{Schonwald}},
  \bibinfo{journal}{Phys. Lett.} \textbf{\bibinfo{volume}{B791}},
  \bibinfo{pages}{206} (\bibinfo{year}{2019}), \eprint{1901.08018}.

\bibitem[{\citenamefont{Blumlein et~al.}(2012)\citenamefont{Blumlein,
  De~Freitas, and van Neerven}}]{Blumlein:2011mi}
\bibinfo{author}{\bibfnamefont{J.}~\bibnamefont{Blumlein}},
  \bibinfo{author}{\bibfnamefont{A.}~\bibnamefont{De~Freitas}},
  \bibnamefont{and} \bibinfo{author}{\bibfnamefont{W.}~\bibnamefont{van
  Neerven}}, \bibinfo{journal}{Nucl. Phys.} \textbf{\bibinfo{volume}{B855}},
  \bibinfo{pages}{508} (\bibinfo{year}{2012}), \eprint{1107.4638}.

\bibitem[{\citenamefont{Yennie et~al.}(1961)\citenamefont{Yennie, Frautschi,
  and Suura}}]{Yennie:1961ad}
\bibinfo{author}{\bibfnamefont{D.~R.} \bibnamefont{Yennie}},
  \bibinfo{author}{\bibfnamefont{S.~C.} \bibnamefont{Frautschi}},
  \bibnamefont{and} \bibinfo{author}{\bibfnamefont{H.}~\bibnamefont{Suura}},
  \bibinfo{journal}{Annals Phys.} \textbf{\bibinfo{volume}{13}},
  \bibinfo{pages}{379} (\bibinfo{year}{1961}).

\bibitem[{\citenamefont{Kuraev and Fadin}(1985)}]{Kuraev:1985hb}
\bibinfo{author}{\bibfnamefont{E.~A.} \bibnamefont{Kuraev}} \bibnamefont{and}
  \bibinfo{author}{\bibfnamefont{V.~S.} \bibnamefont{Fadin}},
  \bibinfo{journal}{Sov. J. Nucl. Phys.} \textbf{\bibinfo{volume}{41}},
  \bibinfo{pages}{466} (\bibinfo{year}{1985}), \bibinfo{note}{[Yad.
  Fiz.41,733(1985)]}.

\bibitem[{\citenamefont{Nicrosini and Trentadue}(1987)}]{Nicrosini:1986sm}
\bibinfo{author}{\bibfnamefont{O.}~\bibnamefont{Nicrosini}} \bibnamefont{and}
  \bibinfo{author}{\bibfnamefont{L.}~\bibnamefont{Trentadue}},
  \bibinfo{journal}{Phys. Lett.} \textbf{\bibinfo{volume}{B196}},
  \bibinfo{pages}{551} (\bibinfo{year}{1987}).

\bibitem[{\citenamefont{Nicrosini and Trentadue}(1988)}]{Nicrosini:1987sw}
\bibinfo{author}{\bibfnamefont{O.}~\bibnamefont{Nicrosini}} \bibnamefont{and}
  \bibinfo{author}{\bibfnamefont{L.}~\bibnamefont{Trentadue}},
  \bibinfo{journal}{Z. Phys.} \textbf{\bibinfo{volume}{C39}},
  \bibinfo{pages}{479} (\bibinfo{year}{1988}).

\bibitem[{\citenamefont{Berends et~al.}(1988)\citenamefont{Berends, van
  Neerven, and Burgers}}]{Berends:1987ab}
\bibinfo{author}{\bibfnamefont{F.~A.} \bibnamefont{Berends}},
  \bibinfo{author}{\bibfnamefont{W.~L.} \bibnamefont{van Neerven}},
  \bibnamefont{and} \bibinfo{author}{\bibfnamefont{G.~J.~H.}
  \bibnamefont{Burgers}}, \bibinfo{journal}{Nucl. Phys.}
  \textbf{\bibinfo{volume}{B297}}, \bibinfo{pages}{429} (\bibinfo{year}{1988}),
  \bibinfo{note}{[Erratum: Nucl. Phys.B304,921(1988)]}.

\bibitem[{\citenamefont{Denner et~al.}(2000)\citenamefont{Denner, Dittmaier,
  Roth, and Wackeroth}}]{Denner:2000bj}
\bibinfo{author}{\bibfnamefont{A.}~\bibnamefont{Denner}},
  \bibinfo{author}{\bibfnamefont{S.}~\bibnamefont{Dittmaier}},
  \bibinfo{author}{\bibfnamefont{M.}~\bibnamefont{Roth}}, \bibnamefont{and}
  \bibinfo{author}{\bibfnamefont{D.}~\bibnamefont{Wackeroth}},
  \bibinfo{journal}{Nucl. Phys.} \textbf{\bibinfo{volume}{B587}},
  \bibinfo{pages}{67} (\bibinfo{year}{2000}), \eprint{hep-ph/0006307}.

\bibitem[{\citenamefont{Cooper-Sarkar et~al.}(2011)\citenamefont{Cooper-Sarkar,
  Mertsch, and Sarkar}}]{CooperSarkar:2011pa}
\bibinfo{author}{\bibfnamefont{A.}~\bibnamefont{Cooper-Sarkar}},
  \bibinfo{author}{\bibfnamefont{P.}~\bibnamefont{Mertsch}}, \bibnamefont{and}
  \bibinfo{author}{\bibfnamefont{S.}~\bibnamefont{Sarkar}},
  \bibinfo{journal}{JHEP} \textbf{\bibinfo{volume}{08}}, \bibinfo{pages}{042}
  (\bibinfo{year}{2011}), \eprint{1106.3723}.

\bibitem[{\citenamefont{Gandhi et~al.}(1998)\citenamefont{Gandhi, Quigg, Reno,
  and Sarcevic}}]{Gandhi:1998ri}
\bibinfo{author}{\bibfnamefont{R.}~\bibnamefont{Gandhi}},
  \bibinfo{author}{\bibfnamefont{C.}~\bibnamefont{Quigg}},
  \bibinfo{author}{\bibfnamefont{M.~H.} \bibnamefont{Reno}}, \bibnamefont{and}
  \bibinfo{author}{\bibfnamefont{I.}~\bibnamefont{Sarcevic}},
  \bibinfo{journal}{Phys. Rev.} \textbf{\bibinfo{volume}{D58}},
  \bibinfo{pages}{093009} (\bibinfo{year}{1998}), \eprint{hep-ph/9807264}.

\bibitem[{\citenamefont{Connolly et~al.}(2011)\citenamefont{Connolly, Thorne,
  and Waters}}]{Connolly:2011vc}
\bibinfo{author}{\bibfnamefont{A.}~\bibnamefont{Connolly}},
  \bibinfo{author}{\bibfnamefont{R.~S.} \bibnamefont{Thorne}},
  \bibnamefont{and} \bibinfo{author}{\bibfnamefont{D.}~\bibnamefont{Waters}},
  \bibinfo{journal}{Phys. Rev.} \textbf{\bibinfo{volume}{D83}},
  \bibinfo{pages}{113009} (\bibinfo{year}{2011}), \eprint{1102.0691}.

\bibitem[{\citenamefont{Albacete et~al.}(2015)\citenamefont{Albacete, Illana,
  and Soto-Ontoso}}]{Albacete:2015zra}
\bibinfo{author}{\bibfnamefont{J.~L.} \bibnamefont{Albacete}},
  \bibinfo{author}{\bibfnamefont{J.~I.} \bibnamefont{Illana}},
  \bibnamefont{and}
  \bibinfo{author}{\bibfnamefont{A.}~\bibnamefont{Soto-Ontoso}},
  \bibinfo{journal}{Phys. Rev.} \textbf{\bibinfo{volume}{D92}},
  \bibinfo{pages}{014027} (\bibinfo{year}{2015}), \eprint{1505.06583}.

\bibitem[{\citenamefont{Ball et~al.}(2018)\citenamefont{Ball, Bertone, Bonvini,
  Marzani, Rojo, and Rottoli}}]{Ball:2017otu}
\bibinfo{author}{\bibfnamefont{R.~D.} \bibnamefont{Ball}},
  \bibinfo{author}{\bibfnamefont{V.}~\bibnamefont{Bertone}},
  \bibinfo{author}{\bibfnamefont{M.}~\bibnamefont{Bonvini}},
  \bibinfo{author}{\bibfnamefont{S.}~\bibnamefont{Marzani}},
  \bibinfo{author}{\bibfnamefont{J.}~\bibnamefont{Rojo}}, \bibnamefont{and}
  \bibinfo{author}{\bibfnamefont{L.}~\bibnamefont{Rottoli}},
  \bibinfo{journal}{Eur. Phys. J.} \textbf{\bibinfo{volume}{C78}},
  \bibinfo{pages}{321} (\bibinfo{year}{2018}), \eprint{1710.05935}.

\bibitem[{\citenamefont{Aaij et~al.}(2017)}]{Aaij:2016jht}
\bibinfo{author}{\bibfnamefont{R.}~\bibnamefont{Aaij}} \bibnamefont{et~al.}
  (\bibinfo{collaboration}{LHCb}), \bibinfo{journal}{JHEP}
  \textbf{\bibinfo{volume}{06}}, \bibinfo{pages}{147} (\bibinfo{year}{2017}),
  \eprint{1610.02230}.

\bibitem[{\citenamefont{Aaij et~al.}(2013)}]{Aaij:2013mga}
\bibinfo{author}{\bibfnamefont{R.}~\bibnamefont{Aaij}} \bibnamefont{et~al.}
  (\bibinfo{collaboration}{LHCb}), \bibinfo{journal}{Nucl.Phys.}
  \textbf{\bibinfo{volume}{B871}}, \bibinfo{pages}{1} (\bibinfo{year}{2013}),
  \eprint{1302.2864}.

\bibitem[{\citenamefont{Aaij et~al.}(2016)}]{Aaij:2015bpa}
\bibinfo{author}{\bibfnamefont{R.}~\bibnamefont{Aaij}} \bibnamefont{et~al.}
  (\bibinfo{collaboration}{LHCb}), \bibinfo{journal}{JHEP}
  \textbf{\bibinfo{volume}{03}}, \bibinfo{pages}{159} (\bibinfo{year}{2016}),
  \bibinfo{note}{[Erratum: JHEP09,013(2016)]}, \eprint{1510.01707}.

\bibitem[{\citenamefont{Bonvini et~al.}(2016)\citenamefont{Bonvini, Marzani,
  and Peraro}}]{Bonvini:2016wki}
\bibinfo{author}{\bibfnamefont{M.}~\bibnamefont{Bonvini}},
  \bibinfo{author}{\bibfnamefont{S.}~\bibnamefont{Marzani}}, \bibnamefont{and}
  \bibinfo{author}{\bibfnamefont{T.}~\bibnamefont{Peraro}},
  \bibinfo{journal}{Eur. Phys. J.} \textbf{\bibinfo{volume}{C76}},
  \bibinfo{pages}{597} (\bibinfo{year}{2016}), \eprint{1607.02153}.

\bibitem[{\citenamefont{Bonvini et~al.}(2017)\citenamefont{Bonvini, Marzani,
  and Muselli}}]{Bonvini:2017ogt}
\bibinfo{author}{\bibfnamefont{M.}~\bibnamefont{Bonvini}},
  \bibinfo{author}{\bibfnamefont{S.}~\bibnamefont{Marzani}}, \bibnamefont{and}
  \bibinfo{author}{\bibfnamefont{C.}~\bibnamefont{Muselli}},
  \bibinfo{journal}{JHEP} \textbf{\bibinfo{volume}{12}}, \bibinfo{pages}{117}
  (\bibinfo{year}{2017}), \eprint{1708.07510}.

\bibitem[{\citenamefont{Bonvini}(2018)}]{Bonvini:2018iwt}
\bibinfo{author}{\bibfnamefont{M.}~\bibnamefont{Bonvini}}
  (\bibinfo{year}{2018}), \eprint{1805.08785}.

\bibitem[{\citenamefont{Abdul~Khalek et~al.}(2019)\citenamefont{Abdul~Khalek,
  Ethier, and Rojo}}]{AbdulKhalek:2019mzd}
\bibinfo{author}{\bibfnamefont{R.}~\bibnamefont{Abdul~Khalek}},
  \bibinfo{author}{\bibfnamefont{J.~J.} \bibnamefont{Ethier}},
  \bibnamefont{and} \bibinfo{author}{\bibfnamefont{J.}~\bibnamefont{Rojo}}
  (\bibinfo{year}{2019}), \eprint{1904.00018}.

\bibitem[{\citenamefont{Barger et~al.}(2017)\citenamefont{Barger, Basso, Gao,
  and Keung}}]{Barge:2016uzn}
\bibinfo{author}{\bibfnamefont{V.}~\bibnamefont{Barger}},
  \bibinfo{author}{\bibfnamefont{E.}~\bibnamefont{Basso}},
  \bibinfo{author}{\bibfnamefont{Y.}~\bibnamefont{Gao}}, \bibnamefont{and}
  \bibinfo{author}{\bibfnamefont{W.-Y.} \bibnamefont{Keung}},
  \bibinfo{journal}{Phys. Rev.} \textbf{\bibinfo{volume}{D95}},
  \bibinfo{pages}{093002} (\bibinfo{year}{2017}), \eprint{1611.00773}.

\bibitem[{\citenamefont{Aartsen et~al.}(2013)}]{Aartsen:2013jla}
\bibinfo{author}{\bibfnamefont{M.~G.} \bibnamefont{Aartsen}}
  \bibnamefont{et~al.} (\bibinfo{collaboration}{IceCube}), in
  \emph{\bibinfo{booktitle}{Proceedings, 33rd International Cosmic Ray
  Conference \\ (ICRC2013): Rio de Janeiro, Brazil, July 2-9, 2013}}
  (\bibinfo{year}{2013}), \eprint{1309.7003}.

\bibitem[{\citenamefont{Lu}(2018)}]{Lu:2017nti}
\bibinfo{author}{\bibfnamefont{L.}~\bibnamefont{Lu}}
  (\bibinfo{collaboration}{IceCube}), \bibinfo{journal}{PoS}
  \textbf{\bibinfo{volume}{ICRC2017}}, \bibinfo{pages}{1002}
  (\bibinfo{year}{2018}).

\bibitem[{\citenamefont{Aartsen et~al.}(2014)}]{Aartsen:2014njl}
\bibinfo{author}{\bibfnamefont{M.~G.} \bibnamefont{Aartsen}}
  \bibnamefont{et~al.} (\bibinfo{collaboration}{IceCube})
  (\bibinfo{year}{2014}), \eprint{1412.5106}.

\bibitem[{\citenamefont{Mikaelian and Zheleznykh}(1980)}]{Mikaelian:1980vd}
\bibinfo{author}{\bibfnamefont{K.~O.} \bibnamefont{Mikaelian}}
  \bibnamefont{and} \bibinfo{author}{\bibfnamefont{I.~M.}
  \bibnamefont{Zheleznykh}}, \bibinfo{journal}{Phys. Rev.}
  \textbf{\bibinfo{volume}{D22}}, \bibinfo{pages}{2122} (\bibinfo{year}{1980}).

\bibitem[{\citenamefont{Frixione and Webber}(2002)}]{Frixione:2002ik}
\bibinfo{author}{\bibfnamefont{S.}~\bibnamefont{Frixione}} \bibnamefont{and}
  \bibinfo{author}{\bibfnamefont{B.~R.} \bibnamefont{Webber}},
  \bibinfo{journal}{JHEP} \textbf{\bibinfo{volume}{0206}}, \bibinfo{pages}{029}
  (\bibinfo{year}{2002}), \eprint{hep-ph/0204244}.

\bibitem[{\citenamefont{Nason}(2004)}]{Nason:2004rx}
\bibinfo{author}{\bibfnamefont{P.}~\bibnamefont{Nason}},
  \bibinfo{journal}{JHEP} \textbf{\bibinfo{volume}{0411}}, \bibinfo{pages}{040}
  (\bibinfo{year}{2004}), \eprint{hep-ph/0409146}.

\bibitem[{\citenamefont{Frixione et~al.}(2007)\citenamefont{Frixione, Nason,
  and Oleari}}]{Frixione:2007vw}
\bibinfo{author}{\bibfnamefont{S.}~\bibnamefont{Frixione}},
  \bibinfo{author}{\bibfnamefont{P.}~\bibnamefont{Nason}}, \bibnamefont{and}
  \bibinfo{author}{\bibfnamefont{C.}~\bibnamefont{Oleari}},
  \bibinfo{journal}{JHEP} \textbf{\bibinfo{volume}{0711}}, \bibinfo{pages}{070}
  (\bibinfo{year}{2007}), \eprint{0709.2092}.

\end{thebibliography}

\appendix
\section{Supplementary material} \label{App:unc}
A breakdown of the various theoretical uncertainties which enter the neutrino-nucleus cross-section presented in Figure~\ref{fig:Incl} is provided. 
As discussed, these uncertainties are obtained by adding in quadrature the 1$\sigma$~CL uncertainties of the free PDFs and nuclear corrections, which are then combined linearly with the uncertainty due to renormalisation scale variation (denoted as ``perturbative"). 
To highlight the impact of the various corrections, the central value of the prediction which is obtained when each of the resonant (positive) or nuclear (negative) corrections are neglected is also shown.

\begin{figure}[ht]
  \begin{center}
    \makebox{\includegraphics[width=1.0\columnwidth]{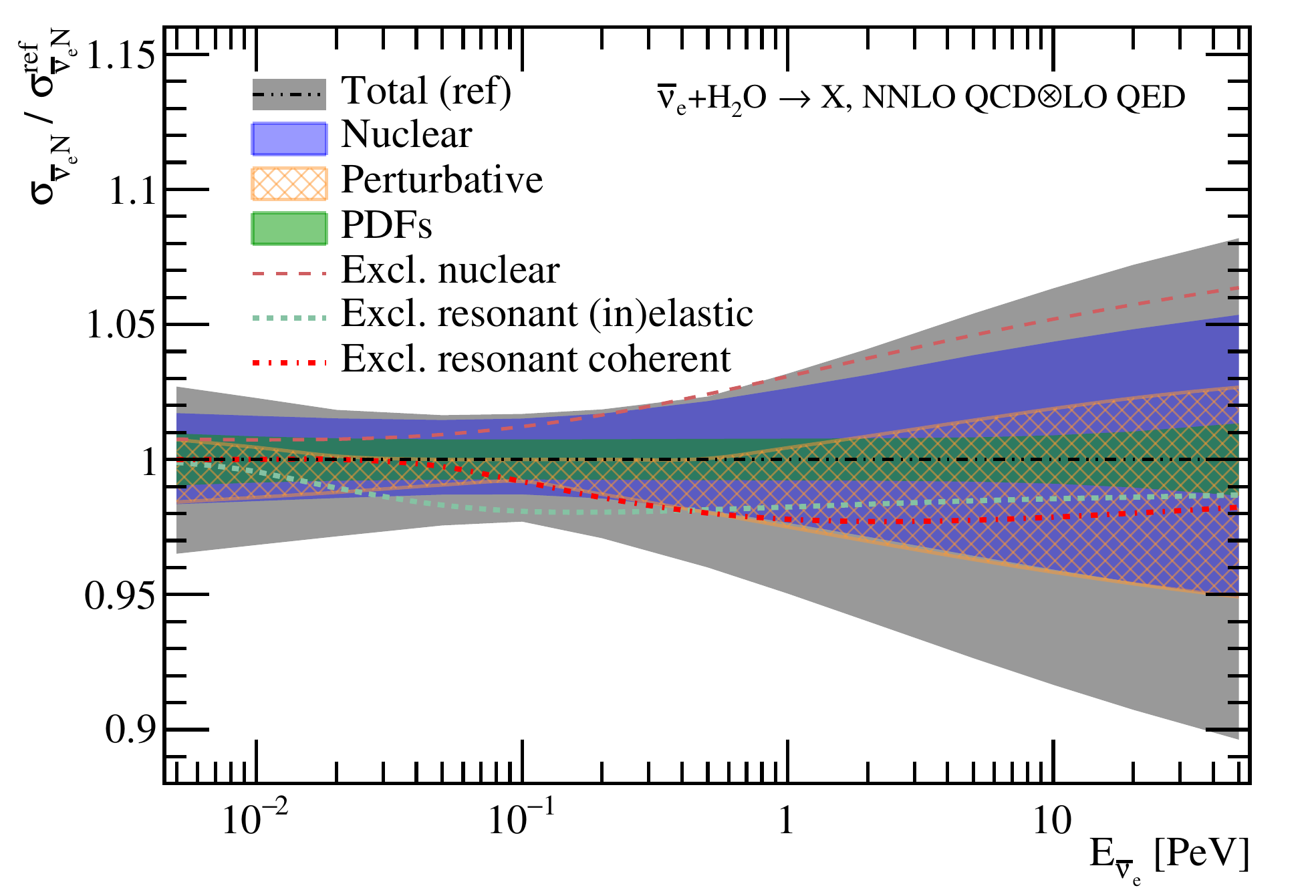}}
  \end{center}
  \vspace{-0.8cm}
  \caption{Breakdown of the various theoretical uncertainties and individual contributions which enter the inclusive cross-section for anti-electron neutrino scattering with an ${\rm H}_{2}{\rm O}$ nucleus.}  
  \label{fig:uncertainties}
\end{figure}

The nuclear corrections (and associated uncertainty) are largest, followed by the perturbative uncertainty of the NNLO calculation which amounts to (3-5)\%. Note that NNLO precision is necessary to reduce the perturbative uncertainty (which is otherwise overwhelming, and typically not assessed) in this calculation. 
These are legitimate uncertainties which should be considered---particularly as DIS measurements in the relevant kinematic regime ($Q^2 \sim m_{\PW}^2$, and $x \sim 10^{-6}$) do not exist, and subprocesses like top-quark production have never been observed in DIS.
In the $\PeV$ range, the resonant contributions accumulatively introduce a positive 4\% correction on the central value. Each of these contributions also have associated uncertainties related to the factorisation scale or the choice of electro magnetic nuclear form factor, for the (in)elastic and coherent contributions respectively. These uncertainties are typically below 20\% (relatively) and therefore have a per mille effect on the total cross-section.

\end{document}